# Correction of FLASH-based MT saturation in human brain for residual bias of $B_1$-inhomogeneity at 3T

Gunther Helms,[1,2]  Nikolaus Weiskopf,[2,3] Antoine Lutti[4]

1 Medical Radiation Physics, Department of Clinical Sciences Lund, Lund University, Lund, Sweden

2 Dept. of Neurophysics, Max-Planck-Institute for Human Cognitive and Brain Sciences, Leipzig, Germany

3 Wellcome Centre for Human Neuroimaging, UCL Institute of Neurology, London, United Kingdom

4 Laboratory for Research in Neuroimaging, Department of Clinical Neuroscience, Lausanne University Hospital and University of Lausanne, Lausanne, Switzerland,

## Abstract

Background:

Magnetization transfer (MT) saturation reflects the additional saturation of the MRI signal imposed by an MT pulse and is largely driven by the saturation of the bound pool. This reduction of the bound polarization by the MT pulse is less efficient than predicted by the differential $B_1^{(+)2}$ law. Thus, $B_1^{(+)}$ inhomogeneities lead to a residual bias in the MT saturation maps. We derive a heuristic correction to reduce this bias for a widely used multi-parameter mapping protocol. at 3T.

Methods:

The amplitude of the MT pulse was varied via the nominal flip angle to mimic variations in $B_1^{(+)}$. The MT saturation's dependence on the actual flip angle features a linear correction term, which was determined separately for gray and white matter.

Results:

The deviation of MT saturation from differential $B_1^{(+)2}$ law is well described by a linear decrease with the actual flip angle of the MT pulse. This decrease showed no significant differences between gray and white matter. Thus, the post hoc correction does not need to take different tissue types into account. Bias-corrected MT saturation maps appeared more symmetric and highlighted highly myelinated tracts.

Discussion:

Our correction involves a calibration that is specific for the MT pulse. While it can also be used to rescale nominal flip angles, different MT pulses and/or protocols will require individual calibration.

Conclusion:

The suggested $B_1^{(+)}$ correction of the MT maps can be applied post hoc using an independently acquired flip angle map.

## Introduction

Magnetization transfer (MT) is a contrast mechanism based on the interaction between the MRI-invisible, "bound" protons of ultra-short $T_2$ (approx. 10 µs) associated with macromolecules and those of rotationally mobile "free" water [1]. In MRI applications, MT is often evoked by interleaving an off-resonance band-selective radio-frequency (RF) pulse with 3D spoiled gradient echo MRI. This "MT pulse" predominantly reduces the longitudinal magnetization $M_{zb}$ of the bound pool, which is then transferred to mobile water ($M_{zf}$). Traditionally, the MT effect is expressed as the percentage attenuation of a reference experiment without MT pulse, the MT ratio (MTR). Since the MTR is derived from two steady state signals, it depends on TR and the local flip angle, and thus on local $T_1$ and inhomogeneity of the RF field $B_1^{(+)}$. In the steady state, however, the shorter $T_1$ and larger MT effect associated with higher macromolecular content will counteract each other [2]. Instead, MT saturation can be expressed as the additional percentage reduction of $M_{zf}$ evoked by a single MT pulse and subsequent transfer during TR [3]. This so-called "MT saturation" can be estimated from an approximate MT-FLASH signal equation using the maps of the apparent $T_1$ (not corrected for local flip angle variations) and signal amplitude $S_0$ [4]. The latter are derived from a conventional dual flip angle experiment, that is, two reference images without MT pulse instead of only one proton density (PD) weighted image for MTR. A multi-echo readout turns these three FLASH acquisitions into a multi-parametric mapping (MPM) technique yielding whole brain maps of $R_1=1/T_1$, $R_2^*$, and MT saturation. Data of 1 mm isotropic resolution can be acquired in less than 20 minutes at 3T [5].

At static field strengths exceeding 1.5T, the RF field varies considerably across the human head. Hence, $B_1^{(+)}$ mapping becomes mandatory to correct for the deviations of flip angle on the estimated $T_1$ values [4]. In the MT saturation maps, fortunately, the effects of flip angle deviations are cancelled out to a large degree, since the off-resonance saturation is governed by square $B_1^{(+)}$ via the absorption lineshape of the bound pool [6]. Thus, brain maps of MT saturation display a uniform contrast and three distinct modes of cerebro-spinal fluid (CSF), grey and white matter (GM, WM) in the histograms without any further corrections [4].

This note addresses the residual effects of the local $B_1^{(+)}$ field on the MT saturation obtained from a widely used 3T MPM protocol [7]. The reduction of $M_{zb}$ renders the partial saturation less efficient than predicted by the square $B_1^{(+)}$ law. Hence an empirical correction term was introduced that decreases linearly with $B_1^{(+)}$ and acts as a multiplicative factor on the MT saturation estimates [8]. This correction term was determined by varying the amplitude of the MT pulse via the nominal flip angle and calibrated separately for WM and GM, but no significant differences were observed. Consequently, the suggested $B_1^{(+)}$ correction can be applied voxel-wise post hoc using an independently acquired flip angle map.

**Theory**

The effect of the MT pulse in the context of an MRI sequence can be described by an additional saturation experienced by the free water $M_{zf}$ [3]. In this very specific sense, the term saturation corresponds to the reduction in longitudinal magnetization during an RF pulse, from $M_z^{start}$ to $M_z^{end}$. The degree of saturation is denoted by a lower case delta

$$\delta = \left(M_z^{start} - M_z^{end}\right)/M_z^{start}.$$

[1]

The signal of the MT FLASH experiment can thus be derived in analogy to a dual excitation FLASH sequence, where the partial saturation of $M_z$ due to the second excitation with flip angle $\alpha_2$, $1 - \cos\alpha_2$, is formally replaced by the "MT saturation" $\delta_{MT}$. Excitation of the water signal and $T_1$ relaxation of the spin system is taken into account as usual [4].

Neglecting exchange during the MT pulse, the differential reduction of $M_z$ in each pool is given by

$$dM_{b/f}/dt = -\pi\, g_{b/f}(\Delta)\, \omega_{1sat}^2\, M_{b/f},$$

[2]

where z has been replaced by f or b to denote the pool. $g_{b/f}(\Delta)$ is the absorption lineshape of the respective pool at frequency offset $\Delta$, and $\omega_{1sat} = \gamma\, B_{1sat}$ the shape of the MT pulse [6]. The free-pool magnetization $M_f$ is hardly changed during the MT pulse at 2 kHz offset, so the tiny saturation $\delta_f$ depends on the squared flip angle $\alpha_{sat}$ (referring to on resonance conditions) of the MT pulse of fixed shape and duration:

$$\delta_f(\alpha_{sat}) = \delta_{0f}\alpha_{sat}^2.$$

[3]

In the bound pool, however, the reduction of $M_b$ during the MT pulse has to be considered when integrating Eq. [2]. This will result in a $\delta_b$ that is increasing with less than the square of $\alpha_{sat}$. Empirically, this behaviour can be modelled by an additional linear factor with an arbitrary constant *q*:

$$\delta_b(\alpha_{sat}) = \delta_{0b}\alpha_{sat}^2(1 - q\alpha_{sat})$$

[4]

These pools, however, cannot be measured directly. The MT saturation $\delta_{MT}$ is observed as an effect on the free pool, where $\delta_f$ is enhanced by MT, and is driven by the difference $\delta_b - \delta_f$ [3]. From equations 3 and 4, the dependence of $\delta_{MT}$ on $\alpha_{sat}$ can thus be described by two empirical parameters *A* and *B*:

$$\delta_{MT}(\alpha_{sat}) = A\alpha_{sat}^2(1 - B\,\alpha_{sat})$$

[5]

The inhomogeneity of $B_1^{(+)}$ is described by the transmit bias field $f_T(x)$ denoting the deviation of the actual flip angle $\alpha(x)$ from the nominal setting $\alpha_{nom}$:

$$\alpha(x) = f_T(x)\alpha_{nom}$$

[6]

The MT saturation, is calculated from the FLASH images using nominal excitation flip angles, and hence referred to as $\delta_{MTapp}$. This corresponds to using the maps of apparent $T_1$ and apparent signal amplitude (not corrected for local flip angle variations) obtained from the two reference scans. Their dependence on $f_T$ then cancels out the $f_T^2$ introduced by the leading $\alpha_{sat}^2$ in Eq. [5] [4]. Hence

$$\delta_{MTapp}(f_T\,\alpha_{sat}) = A\alpha_{sat}^2(1 - Bf_T\,\alpha_{sat}).$$

[7]

For a given MT pulse over a range of actual flip angles, this empirical equation describes the dependence of $\delta_{MTapp}$ on the nominal $\alpha_{sat}$ and local $f_T$. Local *A* and $Bf_T$ are determined by linear regression of $\delta_{MTapp}/\alpha_{sat}^2$ over $\alpha_{sat}$. The $B_1^{(+)}$-corrected MT saturation (that is, without flip angle bias in the linear term) is obtained by substituting the tissue-dependent *A* using Eq. [7]:

$$\delta_{MT}(f_T=1) = A\alpha_{sat}^2(1 - B\alpha_{sat}) = \delta_{MTapp}(f_T)\,(1 - B\alpha_{sat})/(1 - B\alpha_{sat}f_T).$$

[8]

Denoting the product of *B* and nominal $\alpha_{sat}$ as *C*, the correction for the residual influence of $B_1^{(+)}$ inhomogenities on the MT saturation for the given $\alpha_{sat}$ finally takes the form:

$$\delta_{MT} = \delta_{MTapp} \frac{1-C}{1-Cf_T}.$$

[9]

Our results show that *B* and thus *C* do not differ between GM and WM, indicating that this correction can be applied post hoc, irrespective of the underlying tissue class.

## Methods

Experiments were carried out on 5 healthy adults (23 – 28 years, 2 female) on three 3T clinical MRI systems of the same model (Magnetom TIM Trio, Siemens Healthcare, Erlangen, Germany) using the 32 channel head coil or an 8 channel head coil (Invivo, Gainesville, FA) for RF signal reception. Informed written consent according to the local ethical permission was obtained prior to each examination.

To cut down measurement time, the customized multi-gradient-echo MT protocol[5] was modified to 1.25 mm resolution (192x192x128 matrix) and phase and slice directions were encoded with 6/8 partial Fourier acquisition. Thus, the measurement time was approximately 4 min for the $T_1$-weighted and 5.5 minutes for the PD-and MT-weighted datasets. The MT pulse was a 4 ms Gaussian of 220° nominal flip angle and 2 kHz offset. The nominal flip angle was then varied through the values $\alpha_{sat}$ = 90°, 120°, 150°, 180°, 200°, 220°, and 250°. MT maps at lower $\alpha_{sat}$ exhibited a lower SNR, because MT saturation strongly decreases with the power of the MT pulse (Eq. [5]).

Flip angle mapping by the ratio of a stimulated echo and a spin echo was performed as described in [9]. A $T_1$-weighted 3D MP-RAGE (magnetization-prepared rapid-acquisition of gradient echoes) at 1 mm isotropic resolution was acquired to serve as an individual template for co-registration.

Data processing was scripted using the FMRIB software library FSL 6.1 (Centre for Functional MRI of the Brain, Oxford, U.K., www.fmrib.ox.ac.uk/fsl). The FLASH data were averaged across the eight echoes to reduce noise [10] and aligned to the $T_1$-weighted structural volume by a rigid body transformation. MT saturation maps without $B_1^{(+)}$ correction were calculated as described in [4] over a brain mask derived from the PD-weighted dataset. Regions-of-interest (ROIs) were placed in the genu of the corpus callosum, the head of the caudate nucleus, representing WM, and GM, respectively.

The different MT saturation maps were normalized by nominal $\alpha^2_{sat}$ (in radians), and linear regression was performed in each pixel to estimate maps of *A* and slope, which yielded *B* after division by *A* and $f_T$ (Eq. [7]). To evaluate tissue-dependent effects, the MT-map was segmented into GM, WM, and CSF using FSL's advanced segmentation tool (FAST). The histograms of *B* over each tissue class were evaluated using binary masks (thresholded at >0.8). Overlays and histograms were created using fslview.

## Results

The MT saturation in genu and caudate and was plotted over $\alpha_{sat}^2$ (Figure 1) to illustrate the deviation from the $\alpha^2_{sat}$ proportionality due to the decrease of $M_{zb}$ during the MT pulse. Normalization of $\delta_{MTapp}$ by $\alpha_{sat}^2$ revealed an approximately linear decline between 90° and 250° nominal flip angles (Figure 1). Plotted are also points at lower (60°) and higher $\alpha_{sat}$ (280°, just below the SAR limit) which deviated from Eq. [7]. These were excluded from the linear regression.

After accounting for the local flip angle bias, the average slope *B* in five subjects was slightly but not significantly larger in the genu than in the caudate (p>0.05) (Table 1). The differences in the intercept *A* reflect the tissue-specific MT saturation. Figure 1b also illustrates the distance between the intercept and the measured range of $\alpha_{sat}$. Since the linear extrapolation is merely empirical, *A* may not necessarily conform to the actual asymptotic MT saturation for low-power MT pulses. Nevertheless, the residual $B_1^{(+)}$ effect on $\delta_{MTapp}$ should be largely eliminated in *A* (see below). Because the flip angles in deep brain areas tend to be higher than the nominal setting, the apparent $\delta_{MT}$ values in the ROIs were corrected to lower values (Table 1).

Voxel-wise linear regression of $\delta_{MTapp}/\alpha^2_{sat}$ over the nominal $\alpha_{sat}$ (Eq. [7]) yielded maps of the intercept *A* (Figure 2A) showing a high symmetry between the hemispheres and distinct contrast within WM, e.g. highlighting the optic radiation. The *B* maps (after correction for $f_T$) did not show distinct differences between GM and WM (Figure 2B). Since the *B* map was obtained by dividing $f_T$, the relative error of $B_1^{(+)}$ mapping added to that of the regression. The slowly varying systematic deviations in parenchyma may thus reflect a bias in the $f_T$ map. Larger deviations from parenchyma were seen in CSF and the dura. These pixels accounted for the broad tail in the whole brain histogram.

Histogram analysis of *B* over the segmented WM and GM volumes revealed overlapping narrow distributions (not shown). The slightly larger width in GM (full width at half-maximum ~0.035 rad$^{-1}$ vs. ~0.025 rad$^{-1}$) may be explained by partial volumes of CSF and dura. Across the five subjects, the peak of *B = 0.1039* from combined WM and GM (histogram in Figure 2B) was consistent with the ROI values (Table 1). Thus, Eq. [9] was used for the post hoc $B_1^{(+)}$ correction of MT maps to $\alpha_{sat}$ = 220°, with the parameter *C* = 0.4.

Figure 3 shows the apparent MT saturation before and after correction to $\alpha_{sat}$ = 220°, together with a $B_1^{(+)}$ map. Systematic variations in the apparent MT saturation that spatially correlate with $B_1^{(+)}$ as given by the $f_T$ map (bottom row of Fig. 3) are enhanced by the pseudo-color overlay (range between –0.2 p.u. and 2.2 p.u.). The asymmetry in $\delta_{MTapp}$ seen in posterior WM is eliminated by the correction, emphasizing the myelin rich optic radiations (axial), pyramidal (coronal) and transcallosal tracts (sagittal). In line with the typical distribution of $B_1^{(+)}$ inhomogeneity, the MT saturation of cortical GM and subcortical WM was scaled down, while the deep brain regions and brainstem values were scaled up by the correction. Note that the WM mode became considerably narrower by correction. Because the actual flip angles were mostly higher than the nominal value, this subject showed a systematic shift of the GM and WM peaks to higher MT saturation values after correction. Note that the corrected $\delta_{MT}$ maps (middle row of Fig. 3) exhibited less noise than the *A* maps (top row of Fig. 2). As in any linear regression, *A* maps are subject extrapolation of errors in the pixel-wise fitted *B*. In contrast, $\delta_{MT}$ values are corrected from the actual $f_T$ to 1, that is, within the range of measured $\alpha_{sat}$. Any local deviation from the chosen value $C = B\,\alpha_{sat} = 0.4$ will hardly affect the corrected $\delta_{MT}$ maps, while the residual $B_1^{(+)}$ bias is removed.

## Discussion

A heuristic correction of the residual effects of flip angle inhomogeneity on maps of MT saturation in the human brain is presented for a widely used 3T protocol for multi-parametric brain mapping [7]. The underlying model takes into account that an increasing saturation of the bound pool magnetization renders the absorption less efficient than the differential $B_1^2$ dependence (Eq. [2]). Since the dominating $B_1^{(+)2}$ influence is inherently corrected by the calculation [4], the residual effects are much smaller than the $f_T^2$ bias of $T_1$ [11]. At 3T, the typical $B_1^{(+)}$ deviation across the brain is between –20% and +20% . This will impose corrections between –12% and +15% onto $\delta_{MT}$ compared to –36% and +44% applied to $R_1$.

Equations [8] and [9] represent the correction from the actual local flip angle of the MT pulse $f_T\alpha_{sat}$ back to the nominal $\alpha_{sat}$. Thus, it should be noted that the heuristic correction presented in this paper can also be used to correct for a different nominal setting of $\alpha_{sat}$, e.g. after a reduction of the nominal flip angle required to stay within the SAR limit. In this case, also the leading $\alpha_{sat}^2$ has to be taken into account, since only the intrinsic $B_1^{(+)}$ inhomogeneity is eliminated by the algebra of calculating $\delta_{MT}$ .

The MTR or reduction in the steady state signal, is affected by $B_1^{(+)}$ both via the MT pulse and the read-out flip angle. First order corrections of the MTR by the linear deviation of $B_1^{(+)}$ [12,13] are thus purely phenomenological. The main $B_1^{(+)}$ effects on the MTR are inherently coupled to the $T_1$ relaxation term [2]. In contrast, the MT saturation specifically describes the additional reduction of $M_{zf}$ by one MT pulse, that is, during a single TR. The suggested correction model was thus specifically adapted to i) the underlying algebraic framework and ii) the dynamics of $M_{zb}$ during the MT pulse driving the bound pool magnetization into saturation. We made no attempt to model these dynamics, but chose the linear correction term in Eqs. [7-9] as the simplest model covering the typical $B_1^{(+)}$ range at 3T. Increasing the range of $\alpha_{sat}$ may require more complicated models as suggested by Figure 1b. Note also that the Gaussian MT pulse applied in our study was shorter and had a larger offset than the MT pulses implemented by different manufacturers' product sequence. A higher degree of direct saturation will be compatible with the ansatz of our correction (Eq. [5]). As noted above, however, the correction model has to be calibrated for the specific MT-pulse.

The calibration did not reveal significant differences between WM and GM for the parameters of the heuristic correction. This is in line with quantitative MT studies yielding quite similar $T_{2b}$ describing the super-Lorentzian absorption in GM and WM, both in vitro [14,15] and in vivo [16,17]. This reflects that absorption of the invisible pool is governed by super-Lorentzian lineshape of the membrane lipids. In aiming for an easily applicable post hoc correction (like for $R_1 = 1/T_1$) we did not consider smaller effects that depend on the orientation of myelin sheaths relative to $B_0$ [18].

The experiments were performed with an identical acquisition protocol, but on different scanners (Siemens 3T Trio) with different head coils and software versions. No major differences were found. So, the results were pooled and the correction applied in multi-site reproducibility study [7]. Subsequently, it was implemented in the Matlab-based hMRI toolbox (http://hMRI.info), which provides a comprehensive set of tools and processing pipeline for quantitative multi-parametric data [19].

In summary, this note outlines how the observable MT saturation of an off-resonance MT pulse is influenced by $B_1^{(+)}$ inhomogenetity. Our empirical approach may serve as a template to describe other FLASH-based MT protocol, notwithstanding that these may require non-linear models.

## Acknowledgements:

The research leading to these results has received funding from the European Research Council under the European Union's Seventh Framework Programme (FP7/2007-2013) / ERC grant agreement n° 616905.

This project has received funding from the BMBF (01EW1711A & B) in the framework of ERA-NET NEURON.

The Wellcome Centre for Human Neuroimaging is supported by core funding from the Wellcome [203147/Z/16/Z].

GH acknowledges funding by the Swedish Research Council (NT 2014-6193).

AL is supported by the Swiss National Science Foundation (project grant Nr. 320030_184784) and the ROGER DE SPOELBERCH foundation.

**Table 1: Fitted model parameters**

|  | genu[1] | caudate[1] | brain[2] |
|---|---|---|---|
| $A$ [p.u./rad$^2$] | 0.239±0.091 | 0.102±0.005 | n.a. |
| $B$ [1/rad] | 0.1047±0.061 | 0.1001±0.0022 | 0.1039±0.0025 |
| $C$ | 0.402±0.023 | 0.385±0.009 | 0.399±0.010 |
| $\delta_{MTapp}$ [p.u.] | 2.073±0.066 | 0.871±0.022 | n.a. |
| $\delta_{MT}$ [p.u.] | 2.112±0.148 | 0.928±0.054 | n.a. |
| correction | -1.8% | -6.2% | n.a. |

Values given a mean ± standard deviation over five subjects

[1] from localized ROI

[2] from segmented mask of WM and GM (threshold >0.8)

# Figures

**Figure 1: Dependence of MT saturation on the nominal flip angle $\alpha_{sat}$.**

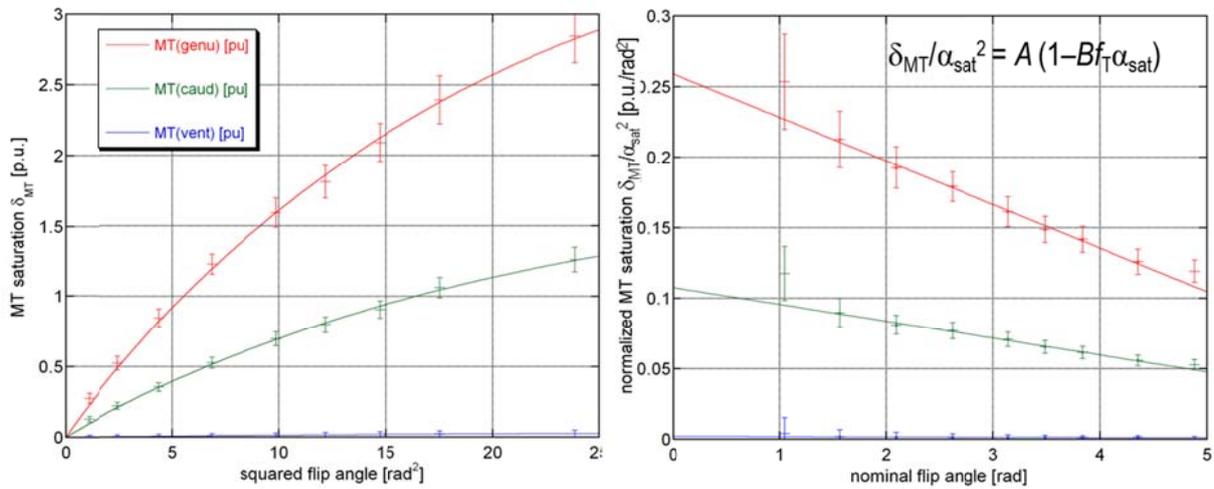

ROI analysis of $\delta_{MTapp}$ in WM (red), GM (green) including additional measurements at 60° and 280° (just below the SAR limit). A ROI in lateral ventricle (blue) shows the behaviour of CSF, where MT is practically absent.

left: Mean ROI values and standard deviations of $\delta_{MTapp}$ are plotted over $\alpha_{sat}^2$ (in radians) to illustrate the increasing deviations from the quadratic dependence.

right: Corresponding values of $\delta_{MTapp}/\alpha_{sat}^2$ over $\alpha_{sat}$ to emphasize the linear correction term (Eq. [5]). The data points at 60° and 280° deviated from linearity and were excluded from regression. Note that the scaling exacerbates noise for decreasing $\alpha_{sat}$

Linear regression yielded in the genu (red) $A = 0.229\pm0.003$ and $Bf_T = 0.112\pm0.003$ ($r = 0.998$) and in the caudate (green) $A = 0.108\pm0.001$ and $Bf_T = 0.111\pm0.002$ ($r = 0.998$). The significant correlation in CSF (blue) [$A = 0.002\pm0.0001$ and $Bf_T = 0.097\pm0.013$ ($r = 0.925$)] was not seen in all subjects but nevertheless deviated strongly from parenchyma.

**Figure 2: Maps and histograms of fitted parameters *A* and *B***

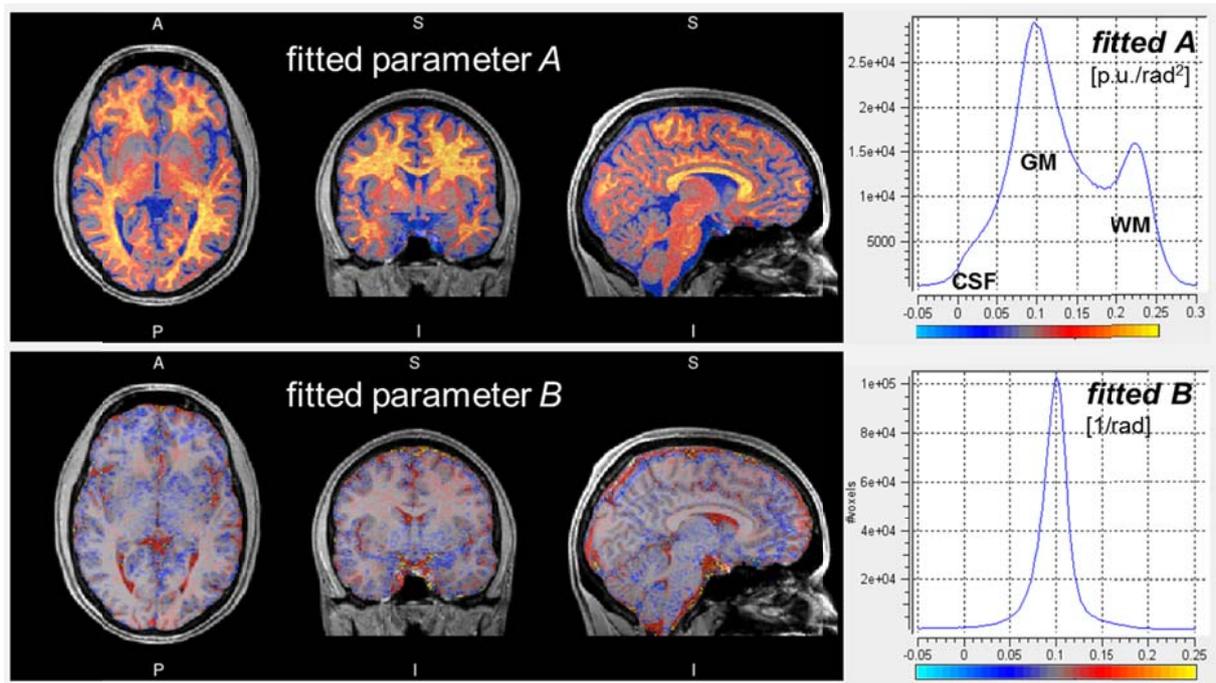

Obtained by pixelwise linear regression (Eq. [7]) and displayed as overlay on MP-RAGE.
Map of fitted *A* (top row): Color-scale centered at *A*=0.1 p.u./rad$^{-2}$ (grey). Highly myelinated tracts (yellow) are discernible against homogeneous WM. The CSF peak is flattened by noise mainly due to extrapolation to $\alpha_{sat} = 0$.
Map of fitted *B* (bottom row): Color-scale centered at *B*=0.1 rad$^{-1}$ (grey). The broad tails of the histogram is due to non-brain pixels (CSF, dura). The CSF peak at 0.12 rad$^{-1}$ (height ~4000 voxels) cannot be discerned in the whole-brain histogram. The distribution of bluish hue may be due to motion or low frequency $B_1^{(+)}$ errors.

**Figure 3: MT saturation maps and histograms prior and after correction**

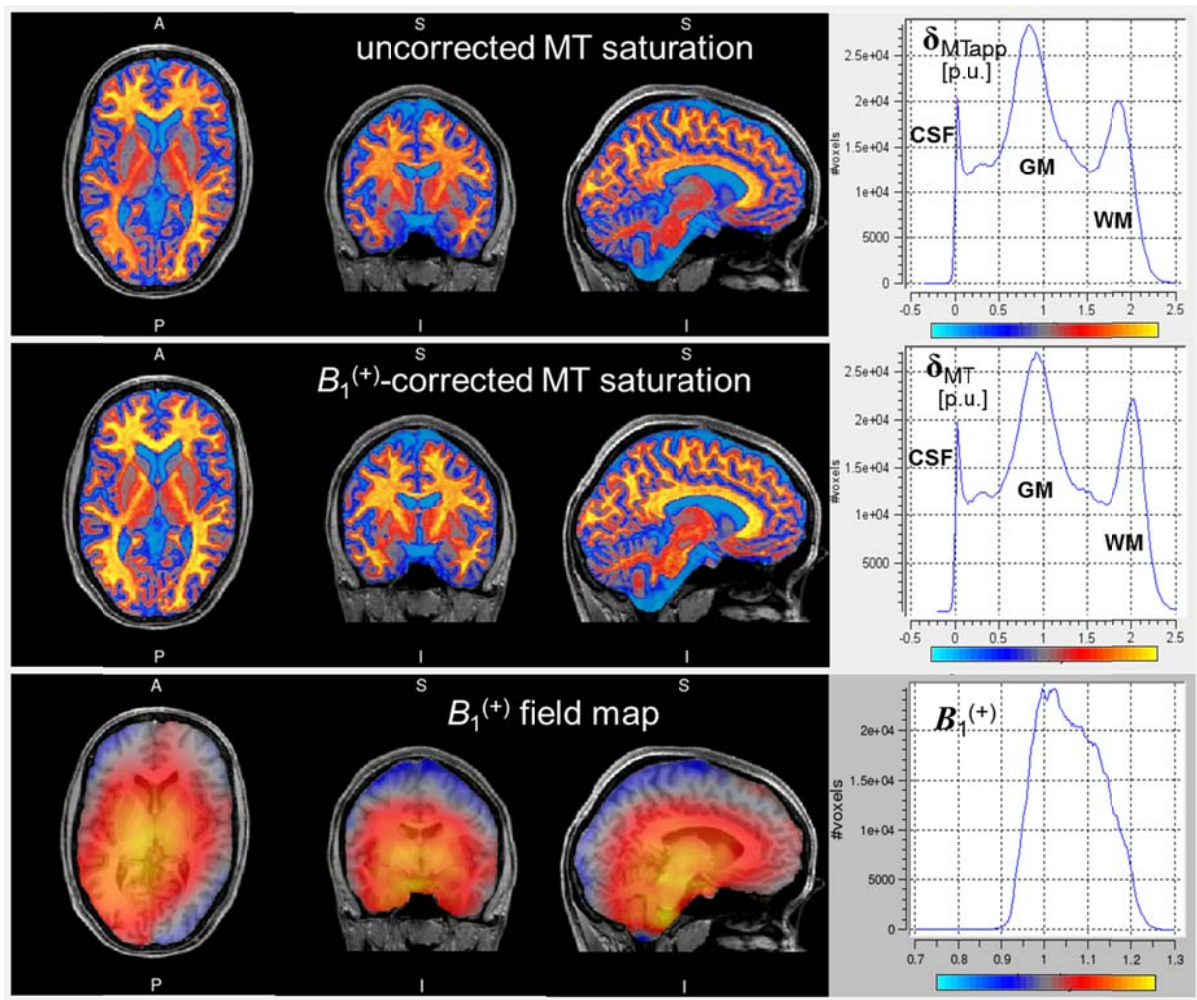

Uncorrected MT saturation map (top row): The color-overlay (center $\delta_{MT}$ = 1 (grey)) enhances the asymmetry in posterior WM.

Corrected MT saturation map (middle row): WM appears more homogeneous with and highly myelinated structures are symmetrically represented by high MT saturation (yellow hue). The WM mode in the histogram is narrowed.

$B_1^{(+)}$ map (bottom row): The color-overlay is centred around $f_T$ = 1 (grey). Underestimation of $\delta_{MTapp}$ is related to higher FA. Since the center of gravity of the histogram is at higher FA, the GM and WM peaks are corrected to higher values.